# SMDP: SARS-CoV-2 Mutation Distribution Profiler for rapid estimation of mutational histories of unusual lineages

Authors: Erin E. Gill[1], Sheri Harari[2], Aijing Feng[3], Fiona S.L. Brinkman[1], Sarah Otto[4]*

Affiliations:
[1] Department of Molecular Biology and Biochemistry, Simon Fraser University, Burnaby, BC, Canada
[2] Fred Hutchinson Cancer Research Center, Howard Hughes Medical Institute, Seattle, WA, USA
[3] Bond Life Sciences Center, University of Missouri, Columbia, MO, USA
[4] Department of Zoology & Biodiversity Research Centre, University of British Columbia, Vancouver, BC, Canada
[*] Corresponding author: otto@zoology.ubc.ca

## Abstract

SARS-CoV-2 usually evolves at a relatively constant rate over time. Occasionally, however, lineages arise with higher-than-expected numbers of mutations given the date of sampling. Such lineages can arise for a variety of reasons, including selection pressures imposed by evolution during a chronic infection or exposure to mutation-inducing drugs like molnupiravir. We have developed an open-source web-based application (SMDP: SARS-CoV-2 Mutation Distribution Profiler; https://eringill.shinyapps.io/covid_mutation_distributions/) that compares a list of user-submitted lineage-defining mutations or a FASTA file containing a single genome (from which lineage-defining mutations are calculated) with established mutation distributions including those observed during (1) the first nine months of the pandemic, (2) during the global transmission of Omicron, (3) during the chronic infection of immunocompromised patients, and (4) during zoonotic spillover from humans to deer. The application calculates the most likely distribution for the user's mutation list and displays log likelihoods for all distributions. In addition, the transition:transversion ratio of the user's list is calculated to determine whether there is evidence of exposure to a mutation-inducing drug such as molnupiravir and indicates whether the list contains mutations in the proofreading domain of nsp14 which could lead to a higher-than-expected mutation rate in the lineage. This tool will be useful for public health and researchers seeking to rapidly infer evolutionary histories of SARS-CoV-2 variants, which can aid risk assessment and public health responses.

## Background

SARS-CoV-2 exhibits a strong clock-like pattern of evolution, with mutational changes at a rate proportional to time, but this pattern is punctuated by “saltational” changes, where lineages suddenly appear with a higher number of mutations than expected given their divergence time from other lineages (Neher, 2022). Such lineages also show unusual distributions of mutations across the genome, often concentrating in spike, and they typically root deep in the SARS-CoV-2 tree, suggesting long periods of time out of circulation. These patterns have been observed in major variants of concern (VoC), such as Alpha, Beta, Gamma, and Delta (Neher, 2022; Otto et al., 2021). and saltational lineages continue to arise (e.g., BA.2.86*, https://github.com/sars-cov-2-variants/lineage-proposals/issues/606, and more recent lineages such as https://github.com/sars-cov-2-variants/lineage-proposals/issues/1584).

Some of these unusual lineages are thought to reflect long passage times within individuals who have weakened immune systems, sharing many of the same signatures seen in chronic infections (Harari et al., 2022). Cele et al. (2022), for example, documented evolutionary changes within a single individual over the course of 190 days that paralleled many of the mutational changes seen in VoC, resulting in heightened immune evasiveness.

When unusual lineages arise, however, it is challenging to infer the evolutionary history leading to the observed genomic changes. Other processes, including passage through animals (Bashor et al., 2021; Kotwa et al., 2023; Naderi et al., 2023), mutator lineages with error-prone polymerases (Takada et al., 2023), and exposure to mutagens such as molnupiravir (Gruber et al., 2024), may also leave unusual genomic signatures.

We developed a shiny application (https://eringill.shinyapps.io/covid_mutation_distributions/) to quantify the strength of the genomic signal in an unusual lineage for different past evolutionary environments. The application requires the user to provide the set of unique mutations leading to the lineage of interest, not including mutations that arose prior to its most recent common ancestor with other lineages or a FASTA file containing a single genome (from which these mutations are calculated using the NextClade Command Line Interface (CLI) (Hadfield et al., 2018)), (see Fig. 1: Data Flow). A sufficient number of mutations is needed to disentangle different genomic signatures, so we recommend the application be used for lineages with at least 10 and preferably ≥20 identifying mutations (see *Power analysis*).

At present, the application compares the similarity of a user-provided list of nucleotide mutations to random draws from the following four distributions:

- A list of mutations observed during the first nine months of the pandemic, prior to the spread of VoC (Harari et al., 2022). (**global pre-VoC distribution**)
- A list of mutations observed during global transmission of Omicron (submission dates up to 25 May 2022), excluding lineage-defining mutations (Harari et al., 2022). (**global Omicron distribution**)
- A list of mutations compiled from 32 chronic infections of immunocompromised individuals (27 individuals from Harari et al., 2022, with 5 additional individuals from Harari et al. 2024) (**chronic distribution**)

- A list of mutations inferred from 109 separate zoonotic spillovers from humans to white-tailed deer (Feng et al., 2023). (**deer distribution**)

The application reports the likelihood of observing the mutation set from each of these four empirically-derived distributions, highlighting which distribution is most likely to explain the observed data and how much more likely that distribution is relative to the next most likely distribution. For these empirically-derived distributions, we consider a distribution that fits the user's mutation set >20 times better than the next best distribution to provide evidence of similarity to that distribution (see *Power analysis*).

Importantly, the application compares only a limited number of empirically-derived distributions from which a given mutation set may be drawn. Results obtained from this application thus describe the similarity only to the given distributions and do not prove that the history of the unusual lineage reflects the most likely distribution, given that other distributions – not examined – may be plausible (see *Discussion*).

While many other animals are susceptible to SARS-CoV-2, there is a very high incidence in white-tailed deer, with active infections observed in 21% of the animals tested in New York State (Feng et al., 2023). A high incidence increases the risk of a reverse spillover, from animals back to humans, as was inferred for three cases in humans (Feng et al., 2023). As more datasets compiling substantial numbers of mutations within other animal species become available, these can be added for further comparisons.

In addition, the application indicates whether the mutation set contains signals consistent with:

- Past molnupiravir use: The transition-to-transversion ratio of mutations is calculated in the focal lineage. For comparison, a background ratio of ~2:1 is typical for SARS-CoV-2, while case-control cohort studies find a higher ratio of ~14:1 under molnupiravir treatment (Gruber et al., 2024). A high ratio thus suggests past exposure to molnupiravir or a similar factor inducing transition mutations.
- Mutator lineages: Mutator alleles may contribute to the unusual features of a lineage by increasing the rate and type of mutation. Known mutators have been observed in nsp14 within the ExoN proofreading domain of SARS-CoV-2. P203L in nsp14 was shown to have an elevated substitution rate in phylogenetic analyses and was confirmed to double the mutation rate when passaged through hamsters (Takada et al., 2023). Sites F60S and C39F in nsp14 were associated with a 22-fold and 6-fold higher substitution rate in phylogenetic analyses (Mack et al., 2023). We considered mutations at sites 39, 60, and 203 in nsp14 to be known mutators and mutations in sites 90, 92, 191, 268, and 273, which fall within the ExoN proofreading domain of nsp14, to be potential mutators (Mack et al., 2023; Moeller et al., 2022).

## Methods

*Distribution of mutations observed globally and in chronic patients:* The data from Table S4 of Harari et al. (2022), as illustrated in their Figures 1 and S1, were filtered to include mutation counts reported as "Chronic infection" to obtain the chronic distribution, "3 Global sequences (12/2019-08/2020)” to obtain the global distribution early in the pandemic (prior to VoC emergence), and “6 Omicron after emergence” to obtain the global distribution late in the pandemic (Omicron era). Mutations in an additional five chronic infections were reported by Harari et al. (2024) and added to the “Chronic infection” set (a total of 32 patients). Mutations were placed in bins consisting of (a) 500 nucleotides, (b) 1000 nucleotides, (c) genes, or (d) genes augmented by splitting spike into pre-RBD, RBD, and post-RBD (receptor binding domain). The latter distribution is given in Table 1.

*Distribution of mutations observed in white-tailed deer:* Aijing Feng and Xiu-Feng Wan (pers. comm.) provided the amino acid identities of mutations that arose in or were associated with white-tailed deer (Feng et al., 2023). Feng et al. inferred the substitutions that had occurred within deer by comparing to the potential human precursor. The distribution of mutations by gene is given in Table 1 and shows a disproportionately high number of mutations in nsp3 (Figure 2).

*FASTA file alignment and determination of lineage-defining mutations:* When a FASTA file is uploaded, it is aligned with all available Nextstrain datasets (Wuhan, BA.2, BA.2.86, XBB) using the NextClade CLI (Hadfield et al., 2018). Each resulting .tsv file is examined, and the dataset with the best alignment score is selected for subsequent analyses. The private nucleotide mutations (reversion substitutions, labeled substitutions and unlabeled substitutions) resulting from the alignment of the FASTA file with this dataset make up the set of lineage-defining mutations.

*Likelihood calculations:* Calculations are performed on either a comma-separated list of nucleotide positions in the SARS-CoV-2 genome where lineage-defining mutations occur for a clade of interest or the set of lineage-defining mutations derived from analysis of an uploaded FASTA file (See Fig. 1). The application determines the likelihood of observing the mutation set as a random draw from each distribution (global pre-VoC, global Omicron, chronic infections, white-tailed deer). The $\log_e$ likelihood of observing the mutation set from each distribution is displayed, as is the relative likelihood of seeing the data from the best fitting distribution relative to the next best fitting.

$\mathrm{Log}_e$ likelihoods were calculated from the multinomial probability of drawing the mutation set from an empirically-based distribution. Given a user-defined bin size (500 nucleotides, 1000 nucleotides, genes, or genes augmented by spike), the probability of observing $k_i$ mutations in the mutation set when there were $n_i$ in the empirical distribution was calculated for each bin *i*. Taking the natural log and summing over bins yields the overall $\log_e$ likelihood:

$$lnL = \sum_i k_i \, ln\left(\frac{n_i + 1}{\sum_j (n_j + 1)}\right)$$

The addition of one to the empirically-derived distribution counts $(n_i + 1)$ ensures that no bins lack data and is expected to have a conservative effect on the results (making all distributions slightly more similar).

*Power analysis:* We assessed the statistical properties of the relative likelihoods of observing the mutation set from the different empirically-derived distributions, estimating the false positive rate and the false negative rate (power). To estimate the false positive rate, we determined how often randomly drawn mutation sets from the global pre-VoC distribution would suggest a misleadingly high probability of being drawn from the chronic or deer distributions (Figure 3A). Using a threshold of 20-fold for rejecting the null hypothesis of the global distribution in favor of another distribution, <1% of 10,000 randomly generated mutation sets yielded a false positive result (e.g., favoring the chronic distribution in 0.46%, 0.55%, 0.31%, 0.11%, 0.01% of replicates for mutation sets of size 5, 10, 20, 40, and 80, respectively). A threshold of 10-fold also yielded reasonable false positive rates (<2%) for the cases tested.

To estimate the false negative rate (power), we determined how often randomly drawn mutation sets from the chronic distribution would yield a strong enough signal to reject the global pre-VoC distribution (Figure 3B). Using a threshold of 20-fold for rejecting the null hypothesis of the global distribution in favor of the chronic distribution, mutation sets with at least 10 mutations led to a rejection of the null hypothesis in the majority of replicates (>50%), with power rising to >80% with at least 20 mutations in the set.

## Application Example

As an example, we consider B.A.2.86, a highly divergent variant that was identified as an unusual lineage on 13 August 2023 by Hynn Spylor (https://github.com/sars-cov-2-variants/lineage-proposals/issues/606). The lineage bears 41 mutational differences from its most recent common ancestor with BA.2 (diverging ~2 years ago) and dispersed rapidly (found in 8 countries by the end of August).

The 41 lineage-defining mutations were listed by Ryan Harris (https://github.com/cov-lineages/pango-designation/issues/2183) as:

{C897A,G3431T,A7842G,C8293T,G8393A,G11042T,C12789T,T13339C,T15756A,A18492G,ins21608,C21711T,G21941T,T22032C,C22208T,A22034G,C22295A,C22353A,A22556G,G22770A,G22895C,T22896A,G22898A,A22910G,C22916T,del23009,G23012A,C23013A,T23018C,T23019C,C23271T,C23423T,A23604G,C24378T,C24990T,C25207T,A26529C,A26610G,C26681T,C26833T,C28958A}

where we have excluded A12160G, considered in the thread to be a rooting problem. The results shown by the application's analysis of the above mutation set are illustrated in Figure 1. The results indicate that this unusually divergent BA.2.86 variant most likely evolved as a result of a chronic infection, among the distributions compared here.

## Discussion

Understanding the recent history of a virus, especially a genetically unusual virus, can help us to reduce future risks: identifying and clearing chronic infections globally, reducing the risks of treatments that are mutagens, and minimizing risks of spillover events. This application attempts to rapidly aid the interpretation of newly appearing, unusual SARS-CoV-2 lineages by quantifying similarity to signals left by past selection and mutation pressures.

There are several caveats to be kept in mind when interpreting the results. Most importantly, interpreting the relative probability of explaining the observed mutation set is limited to the set of distributions compared. Just because the lineage-defining mutations of an unusual lineage is more consistent with the distribution of mutations observed in chronically infected individuals than the global distribution does not prove that the lineage passaged through chronically infected individuals, given that not all possible distributions are included.

In particular, we lack distributions of evolutionary changes from global transmission events in the recent past (the Omicron-era sequences analysed by Harari et al., included submission dates only up to 25 May 2022), which might provide a more relevant comparison for the unusual lineages appearing today. Because the application requires sufficient depth of information for mutations throughout the genome, the many other animal species that could be reservoirs (Chakraborty et al., 2024) are also not included.

There are additional pieces of information that could be used to infer the evolutionary history of SARS-CoV-2 lineages. We have not used information about the rate of mutations per unit time, whether the mutations are silent or coding, or the specific nature of the amino acid changes induced.

If users have unaligned SARS-CoV-2 genome sequences that they would like to use as input for this tool, they must first place them into a phylogeny to detect lineage-defining mutations. To get started, helpful tools can be found on the UCSC SARS-CoV-2 Genome Browser (https://genome.ucsc.edu/goldenPath/help/covidBrowserIntro.html#data). Users can also visualize and export mutations relative to the founder of the clade that a sequence belongs to using NextClade (https://clades.nextstrain.org/).

Nevertheless, we intend the application to be used as a quick means of categorizing the unusual features in a set of mutations that define a SARS-CoV-2 lineage. Additional mutation distributions could be added in the future.

### Data availability

The *Mathematica* code used to process the data and generate Figure 2 and the Python code to produce the shiny application and Figure 1 are available on Zenodo (DOI:10.5281/zenodo.12735051).

### Acknowledgements

We thank all researchers for their investigations of mutation distributions, which enabled development of this SMDP application. SMDP was developed by members of the Computational Analysis, Modelling and Evolutionary Outcomes (CAMEO https://covarrnet.ca/computational-analysis-modelling-and-evolutionary-outcomes-cameo/) pillar of Canada's Coronavirus Variants Rapid Response Network (CoVaRR-Net https://covarrnet.ca/), with further input by other members of this CAMEO group and the Brinkman Lab. CAMEO is supported by Genome Canada, Innovation, Science and Economic Development Canada (ISED) and the Canadian Institutes of Health Research (CIHR) (grant #ARR-175622). This project was supported by funding from a CoVaRR-Net Rapid Response Research Grant. FSLB is an SFU Distinguished Professor.

**Table 1: Distributions compared.** For each genetic region, the start and end positions were based on NCBI Reference Sequence NC_045512.2, with the receptor binding domain of spike spanning amino acid positions 331-524 as given by (Tai et al., 2020). The distributions of mutation counts are from transmissions within the global data set, either early in the pandemic or late in the Omicron era (Harari et al., 2022), chronic infections (Harari et al., 2022), or white-tailed deer (Feng et al., 2023).

| Genomic segment | Start | End* | Global [early] | Global [late] | Chronic | Deer |
|---|---|---|---|---|---|---|
| nsp1 | 266 | 805 | 23 | 40 | 9 | 19 |
| nsp2 | 806 | 2719 | 59 | 152 | 14 | 58 |
| nsp3 | 2720 | 8554 | 148 | 439 | 47 | 204 |
| nsp4 | 8555 | 10054 | 34 | 87 | 11 | 56 |
| proteinase | 10055 | 10972 | 27 | 75 | 4 | 20 |
| nsp6 | 10973 | 11842 | 26 | 66 | 7 | 34 |
| nsp7 | 11843 | 12091 | 4 | 14 | 2 | 9 |
| nsp8 | 12092 | 12685 | 12 | 30 | 4 | 22 |
| nsp9 | 12686 | 13024 | 11 | 26 | 1 | 4 |
| nsp10 | 13025 | 13441 | 6 | 20 | 0 | 6 |
| polymerase | 13442 | 16236 | 68 | 143 | 19 | 34 |
| helicase | 16237 | 18039 | 35 | 104 | 12 | 21 |
| nsp14 (exonuclease) | 18040 | 19620 | 35 | 99 | 11 | 30 |
| endoRNAse | 19621 | 20658 | 29 | 36 | 6 | 49 |
| methyltransferase | 20659 | 21552 | 19 | 48 | 4 | 18 |
| S (NTD) | 21563 | 22552 | 43 | 117 | 42 | 63 |
| S (RBD) | 22553 | 23134 | 13 | 148 | 44 | 7 |
| S (post RBD) | 23135 | 25384 | 52 | 156 | 20 | 57 |
| ORF3a | 25393 | 26220 | 66 | 106 | 6 | 46 |
| E | 26245 | 26472 | 6 | 16 | 13 | 2 |
| M | 26523 | 27191 | 18 | 82 | 8 | 10 |
| ORF6 | 27202 | 27387 | 9 | 23 | 3 | 11 |
| ORF7a | 27394 | 27759 | 15 | 43 | 8 | 30 |
| ORF7b | 27756 | 27887 | 9 | 22 | 2 | 7 |
| ORF8 | 27894 | 28259 | 22 | 28 | 7 | 30 |
| N | 28274 | 29533 | 83 | 153 | 12 | 51 |
| ORF10 | 29558 | 29674 | 6 | 20 | 0 | 0 |
| **Total** | | | **930** | **2398** | **316** | **898** |

* The end of the genic region is set to the nucleotide just before the next genic region, ensuring that all nucleotides from 266-29674 are counted once, regardless of the end of the coding region.

**Figure 1: Data flow.** SMDP accepts input in the form of a list of lineage-defining mutations or a FASTA file containing a single genome. If the user uploads a FASTA file, the NextClade CLI (Hadfield et al., 2018) will first perform *ad hoc* sequence alignments against all available SARS-CoV-2 Nextstrain datasets (Wuhan, BA.2, BA.2.86, XBB). SMDP will then choose the dataset with the best alignment score for subsequent analyses. A list of lineage-defining mutations (called private nucleotide mutations in NextClade) is calculated based on this dataset. Either the user's uploaded list of lineage-defining mutations or the list of mutations calculated by NextClade are used to 1) calculate the transition:transversion ratio, 2) calculate mutator site changes and 3) determine best-fit and second best-fit mutation distributions (after splitting into user-defined bins and log likelihood calculations).

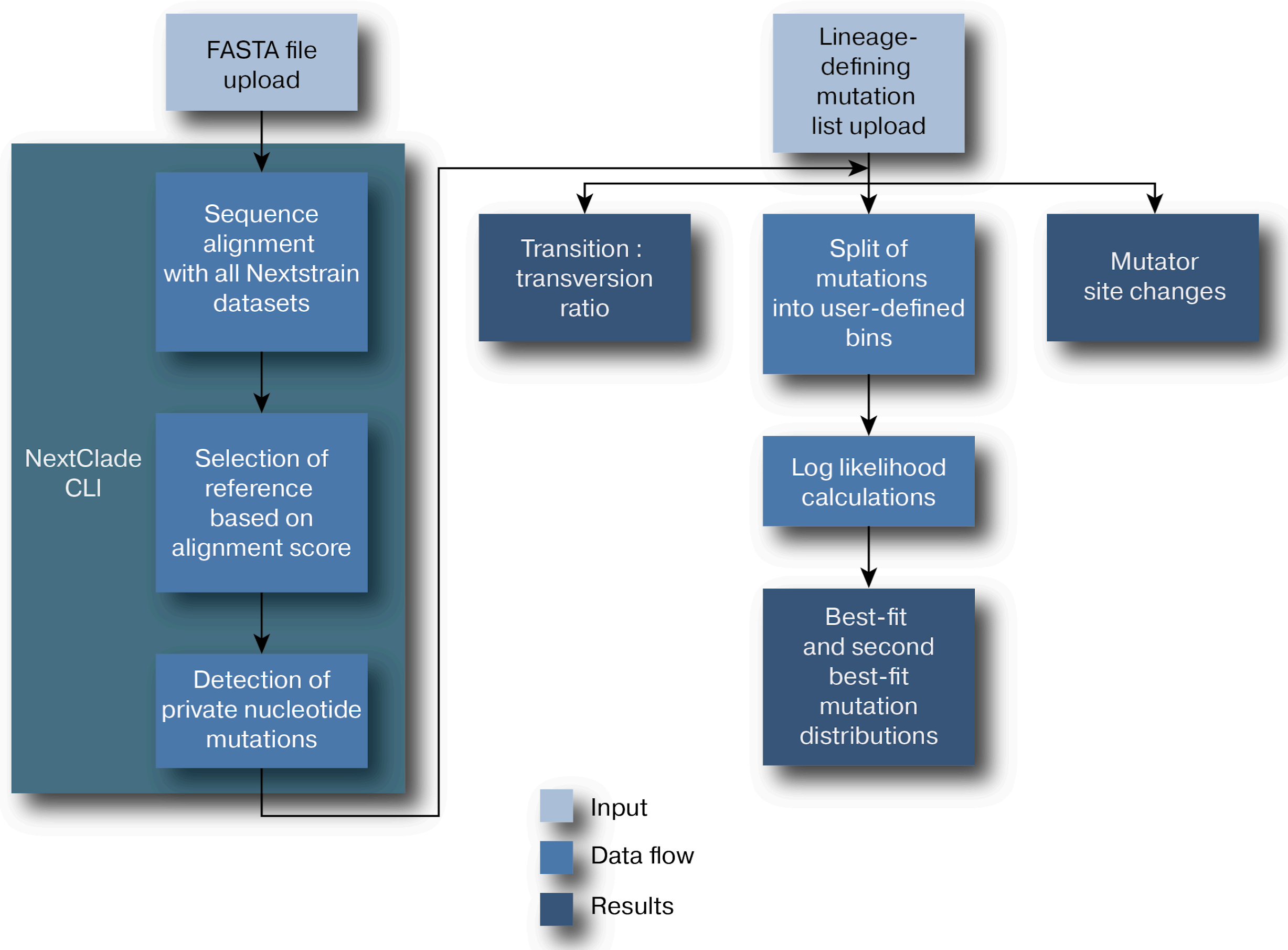

**Figure 2: Usage example.** The lineage-defining mutations in BA.2.86 were used to illustrate the application. The BA.2.86 lineage showed a typical ratio of transitions to transversions (Gruber et al., 2024) and did not carry changes at known mutator sites (Takada et al., 2023). The lineage-defining mutations better matched the chronic distribution than the other distributions (lnL values of -107.51 for the chronic distribution, -137.50 for the global [early] distribution, -116.96 for the global [late] distribution, -144.42 from the deer distribution), being ~13000 times more likely to be drawn from the chronic distribution than the the late global distribution from the Omicron era, the next best matching distribution.

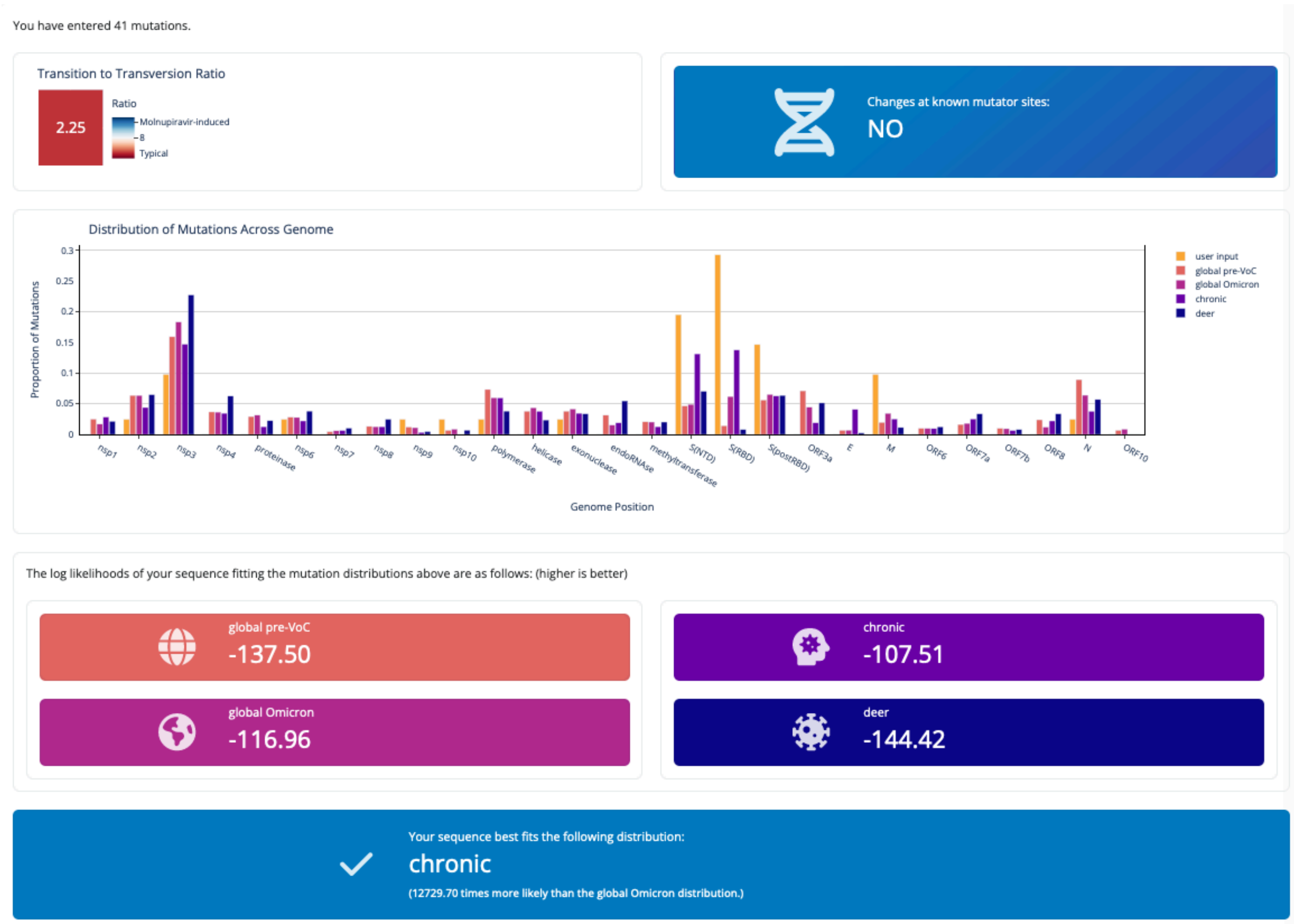

You have entered 41 mutations.
Transition to Transversion Ratio
2.25
Ratio
Molnupiravir-induced
8
Typical
Changes at known mutator sites:
NO
Distribution of Mutations Across Genome
Proportion of Mutations
Genome Position
nsp1
nsp2
nsp3
nsp4
proteinase
nsp6
nsp7
nsp8
nsp9
nsp10
polymerase
helicase
exonuclease
endoRNAse
methyltransferase
S(NTD)
S(RBD)
S(postRBD)
ORF3a
E
M
ORF6
ORF7a
ORF7b
ORF8
N
ORF10
user input
global pre-VoC
global Omicron
chronic
deer
The log likelihoods of your sequence fitting the mutation distributions above are as follows: (higher is better)
global pre-VoC
-137.50
chronic
-107.51
global Omicron
-116.96
deer
-144.42
Your sequence best fits the following distribution:
chronic
(12729.70 times more likely than the global Omicron distribution.)

**Figure 3: Power analysis.** Mutation sets of different sizes (x axis) were drawn at random from either (A) the distribution of global mutations (pre-VoC) or (B) the distribution of chronic mutations (Harari et al., 2022). The null hypothesis that the set is drawn from the global distribution is then tested against the alternative hypothesis that it comes from the chronic (green) or deer (brown) distribution. Relative probabilities (y axis) greater than 20 (above grey shaded region) are rarely seen when the null hypothesis is true (panel A), but become increasingly likely as the size of the mutation set increases when the null hypothesis is false (panel B). Each distribution is obtained from 10,000 replicate mutation sets, with the median (circle), 25-75% quantiles (wide bar), and 2.5-97.5% quantiles (thin bar) illustrated.

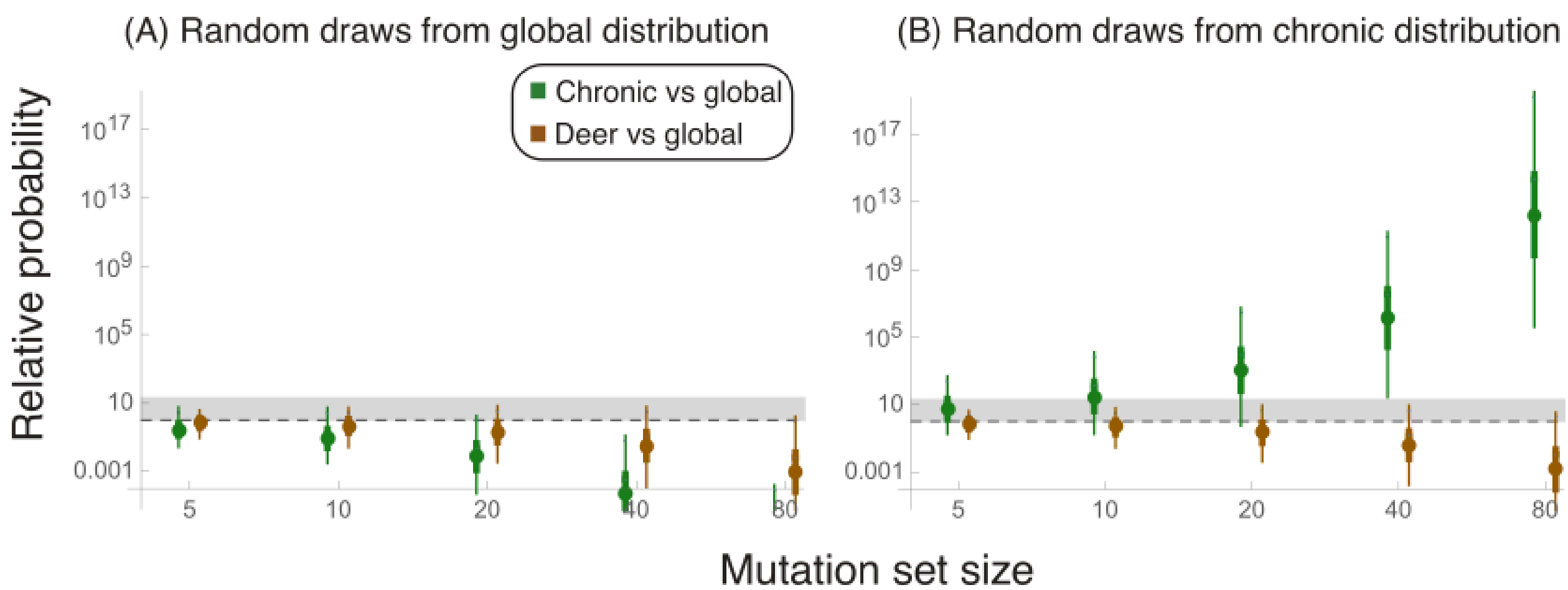

**Table S1: Known and potential mutator sites**. Known sites have been confirmed experimentally, and the specific amino acid / nucleotide changes leading to mutator phenotypes are shown (“Confirmed” in “Site Type” column). Potential sites lie within the ExoN proofreading domain of Nsp14, as shown in Mack et al. (2023) (“Potential” in “Site Type” column). The wild type amino acids and positions within the mature Nsp14 protein, as well as the genomic locations of the encoding nucleotides, are shown for these sites, but changes that would lead to mutator phenotypes have not been confirmed.

| Gene | Amino Acid Change | Nucleotide Change | Site Type | Reference |
|---|---|---|---|---|
| nsp14 | C39F | G18155T | Confirmed | (Mack et al. 2023) |
| nsp14 | F60S | T18218C | Confirmed | (Takada et al. 2023) |
| nsp14 | P203L | C18647T | Confirmed | (Mack et al. 2023) |
| nsp14 | D90 | 18307-18309 (GAT) | Potential | (Mack et al. 2023) |
| nsp14 | E92 | 18313-18315 (GAG) | Potential | (Mack et al. 2023) |
| nsp14 | E191 | 18610-18612 (GAG) | Potential | (Mack et al. 2023) |
| nsp14 | H268 | 18841-18843 (CAT) | Potential | (Mack et al. 2023) |
| nsp14 | D273 | 18856-18858 (GAT) | Potential | (Mack et al. 2023) |